\documentclass[prl,notitlepage,twocolumn]{revtex4}

\usepackage{amsmath}
\usepackage{amssymb}
\usepackage{bm}
\usepackage{graphicx}
\usepackage{times}
\usepackage[outdir=./]{epstopdf}

\usepackage{color}
\definecolor{darkblue}{rgb}{0,0,0.75}

\begin{document}

\title{Metaphotonics with subwavelength dielectric resonators}

\author{Mikhail~V.~Rybin${}^{1,2}$}
\author{Yuri Kivshar${}^{3}$}
\email{yuri.kivshar@anu.edu.au}

\affiliation{$^1$School of Physics and Engineering, ITMO University, St Petersburg 197101, Russia}
\affiliation{$^2$Ioffe Institute, St Petersburg 194021, Russia}
\affiliation{$^3$Nonlinear Physics Center, Research School of Physics, Australian National University, Canberra ACT 2601, Australia\\
$^*$Corresponding author:  yuri.kivshar@anu.edu.au}

\begin{abstract}
The recently emerged \emph{Mie resonant metaphotonics} (or \emph{Mietronics}) provides novel opportunities for subwavelength optics. Mietronics employs resonances in isolated nanoparticles and structured surfaces. We present a brief summary of the key concepts underpinning this rapidly developing area of research, using the examples of isolated high-index dielectric subwavelength particles. We also discuss recent advances and future trends in designs of high-$Q$ elements for efficient resonant spatial and temporal control of light.
\end{abstract}

\maketitle

\section{introduction}

Mie-resonant semiconductor nanoparticles underpin all-dielectric resonant metaphotonics, also termed as 'Mie-tronics' \cite{kivshar2022rise}. Relatively high values of the dielectric refractive index and low absorption losses are associated with a variety of structures ranging from isolated resonators to metasurfaces composed of resonant nanoparticles, and many of their functionalities are based on resonant effects at the nanoscale. Low-order Mie resonances correspond to moderate values of the quality factor ($Q$ factor) up to several tens. Recently, being inspired by the physics of bound states in the continuum (BIC), researchers found a way to increase $Q$ factors of a single nanoparticle~\cite{rybin2017high}, creating a new research direction in subwavelength photonics. The primary goal of this Perspective is to discuss the recent advances in Mie-resonant metaphotonics starting from the key concepts underpinning this active area of research, and then summarizing some applications and the recent trends in all-dielectric resonant photonics. 

\begin{figure*}
\includegraphics{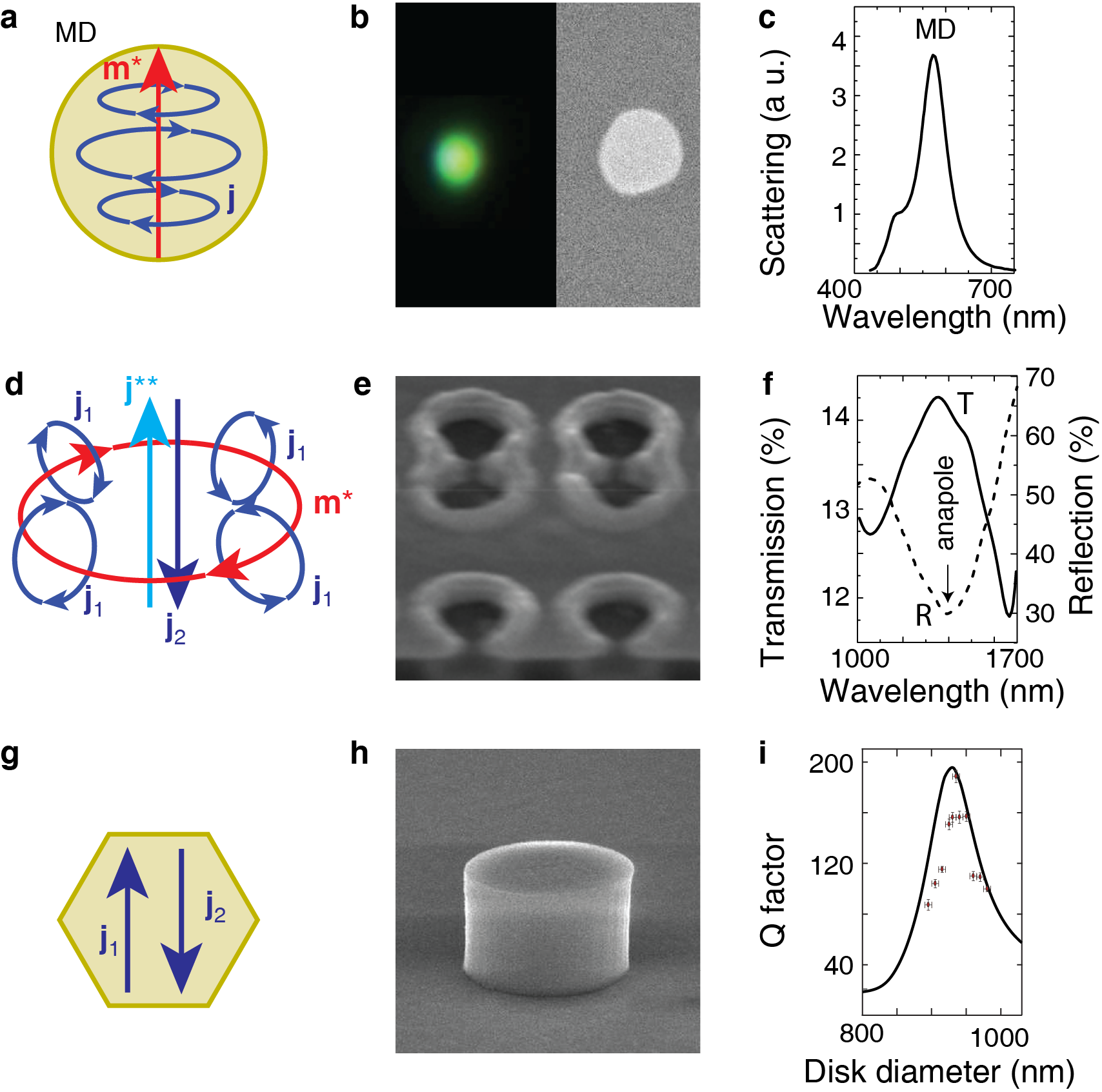}
\caption{\label{fig:concepts}
{\bf Resonances in subwavelength particles.} (a-c) Mie resonances in dielectric particles: \textbf{a} schematic of a magnetic dipole \textbf{m}* formed by coiled displacement currents \textbf{j}; \textbf{b} optical and SEM image of a resonant silicon nanoparticle; \textbf{c} scattering spectra of a nanoparticle shown in \textbf{b}. (d-f) Concept of optical anapole: \textbf{d} Schematic of a toroidal distribution of the displacement currents \textbf{j}$_1$ producing an effective magnetic current coil \textbf{m}* resulting in an effective electric dipole moment \textbf{j}**, which is oscillates out of phase with the real electric dipole moment \textbf{j}$_2$; \textbf{e} SEM image of the structure composed of elements supporting anapole-type distribution of fields; \textbf{f} transmission (solid curve) and reflection (dashed curve) spectra of the structure shown in \textbf{e}. (g-i) Friedrich-Wintgen BIC via interfering resonances: \textbf{g} schematic of two decoupled modes associated with displacement currents \textbf{j}$_1$ and \textbf{j}$_2$ generating field oscillating out of phase in the far field zone; \textbf{h} SEM image of a nanodisk supporting quasi-BIC mode; \textbf{i} $Q$ factor of a nanodisk supporting the quasi-BIC mode. \textbf{b} and \textbf{c} are adapted from \cite{kuznetsov2012magnetic}; \textbf{e} and \textbf{f} are adapted from \cite{wu2018optical}; \textbf{h} and \textbf{i} are adapted from \cite{koshelev2020subwavelength}.}
\end{figure*}

The problem of electromagnetic wave scattering from a spherical particle was solved analytically by Gustav Mie almost a century ago~\cite{bohren1998absorption}. The underlying idea is to exploit the isotropic symmetry of the space and the field representations with the spherical functions to expand the solutions into a series of orthogonal {\it multipoles}. In reality, either a substrate or a non-spherical shape of the particle reduces the symmetry, so the eigenmodes become perturbed and mixed. However, it is still convenient to consider low-order resonances in a subwavelength particle of an arbitrary shape as Mie-like modes (such as {\it electric and magnetic dipoles}, {\it quadrupoles}, etc). Unlike spherical plasmonic subwavelength particles, dielectric particles may support both electric and magnetic modes of the similar strength. Moreover, the magnetic dipole resonance in a spherical particle has the lowest frequency (strictly speaking, for an isotropic and homogeneous case). Although there exist no magnetic currents in nonmagnetic dielectric particles, the effective magnetic dipole moment is induced by circulating displacement currents (see Fig.~\ref{fig:concepts}a, and also 
Refs.~\cite{evlyukhin2010,kuznetsov2012magnetic,wu2018optical,koshelev2020subwavelength}). 

Theoretically, the possibility of optically-induced magnetic resonances in silicon particles in the visible range was suggested by Evlyukhin {\it et al.}~\cite{evlyukhin2010}, 
and the experimental demonstration were reported in 2012 by two groups~\cite{evlyukhin2012,kuznetsov2012magnetic}.  In experiment~\cite{kuznetsov2012magnetic}, the magnetic Mie resonance in a silicon nanosphere manifests itself as a bright spot in a dark-field microscopic image (Fig.~\ref{fig:concepts}b) that provides a peak in the scattering (Fig.~\ref{fig:concepts}c). Despite the existence of nonvanishing absorption in silicon in the visible frequency range, the magnetic dipole resonances are registered even around 500 nm and definitely at longer wavelengths, with the $Q$ factor about 10.

The important concept for wave localization is \emph{destructive interference} in the external or outer space, and this mechanism may improve the $Q$ factor of the Mie resonances. One of the realizations of this idea is associated with the concept of {\it anapole} that was applied to an isolated dielectric nanoparticle in Ref.~\cite{miroshnichenko2015nonradiating}.  
The term 'anapole' was introduced by  Zeldovich in a different context~\cite{zel1957electromagnetic}, but later it was adopted in optics~\cite{papasimakis2016electromagnetic}. The underlying idea of {\it an optical anapole} is illustrated in Fig.~\ref{fig:concepts}d.  If we consider electric currents coiled in a toroid, such a configuration will induce an effective magnetic current along the loop inside the toroid. In turn, the effective magnetic current induces an effective electric dipole (that is referred to as {\it a toroidal dipole}) oscillating along the toroid axis. Now if the system supports a real electric dipole oscillating in the same direction,  the dipole and toroidal modes produce identical far-field radiations, which can be modified by mutual interference. Under certain conditions, destructive interference occurs resulting in the far-filed radiation cancellation. Anapole-based structures were designed for both microwave and optical frequencies (Figs.~\ref{fig:concepts}e,f). In particular, anapole-type scattering cancellation takes place in a high-index dielectric disk~\cite{miroshnichenko2015nonradiating}, but the overall anapole feature depends on the parameters of an incident wave, since anapole is not associated with a specific eigenmode but originating from interference of the modes~\cite{monticone2019can}.

Finally, we wish to mention {\it a bound state in the continuum} (BIC) as one more important concept to  improve the $Q$ factors of isolated dielectric resonators.  Unlike periodic and extended optical structures, in this case BIC is based on interfering resonances. Friedrich and Wintgen~\cite{friedrich1985interfering} examined two quantum states with nonzero radiation decay being coupled only via the near-field term of the radiation. The hybrid modes occur and at a certain parameters each mode produces far-field radiation out of phase with the other one, so the radiation decay vanishes. To some extent, the concept generalizes the anapole state suggested by Zeldovich but with no specification of two eigenmodes. Rybin {\it et al.}~\cite{rybin2017high} adapted the Friedrich-Wintgen concept to optics predicting high-$Q$ resonances for a single dielectric Mie resonator. If a subwavelength particle supports two orthogonal modes, the modes may undergo hybridization due to the near-field coupling resulting in the formation of quasi-BIC states (see Fig.~\ref{fig:concepts}g). For the case of a dielectric disk, the realization of such a quasi-BIC state was termed 'a supercavity mode'. The circular boundary of the disk is responsible for Mie-type modes, and the flat caps confine the wave in the Fabry-Perot-type modes. By varying disk height-to-radius aspect ratio, it is possible to match the frequency of the modes of two types. As a result, strong anti-crossing coupling is observed in the spectra revealing the Friedrich-Wintgen hybridization. The mode dispersion can be described as two branches of low-$Q$ and high-$Q$ modes. The latter has the $Q$ factor depending on the disk aspect ratio, and it undergoes a rapid growth in the vicinity of a certain parameter, being a fingerprint of a quasi-BIC state~\cite{rybin2017supercavity}. Interestingly enough, the mode hybridization leads to other effects, for example, combining two coupled modes allows the so-called superscattering phenomena~\cite{canos2023superscattering}.
It is worth noting that in 2000s a similar concept was employed to increase the $Q$ factors of microcavities~\cite{wiersig2006formation, song2010improving, cao2015dielectric, huang2023resonant}. A microcavity may support a pair of modes coupled outside a resonator resulting in the Friedrich-Wintgen mechanism of enhancement of the $Q$ factors of one of the hybrid modes.

The regime of a supercavity mode can be achieved in other types of high-index dielectric resonators, including a nonspherical dielectric cavity with a rectangular cross section~\cite{mirosh}. In this case, the Mie resonance is replaced by the second  Fabry-Perot resonance,
and high-$Q$ modes are achieved by realizing an avoided crossing of the eigenvalues for a pair of leaky modes. In general, a symmetry analysis makes it possible to determine single-particle modes, which can be employed for the hybridization into a supercavity mode~\cite{gladyshev2020symmetry}.

The existence of a supercavity mode in the water cylinder was experimentally confirmed by Bogdanov {\it et al.}~\cite{bogdanov2019bound}. The aspect ratio was varied by adding water in a circular tube, which makes a  dielectric cylinder. In the microwave range, water has permittivity as high as 80, but tangent loss is considerable as well. Odit {\it et al.}~\cite{odit2021observation} demonstrated a supercavity mode in the microwave regime for a sample made of low-loss ceramics, which enables to trace a sharp quasi-BIC-type $Q$-factor profile as a function of cylinder height. The first observation of the quasi-BIC mode in the optical regime was reported for a AlGaAs nanodisk with $Q$ factor of almost 200 (Figs.~\ref{fig:concepts}h and~\ref{fig:concepts}i) (see Ref.~\cite{koshelev2020subwavelength}). 

\begin{figure*}
\includegraphics{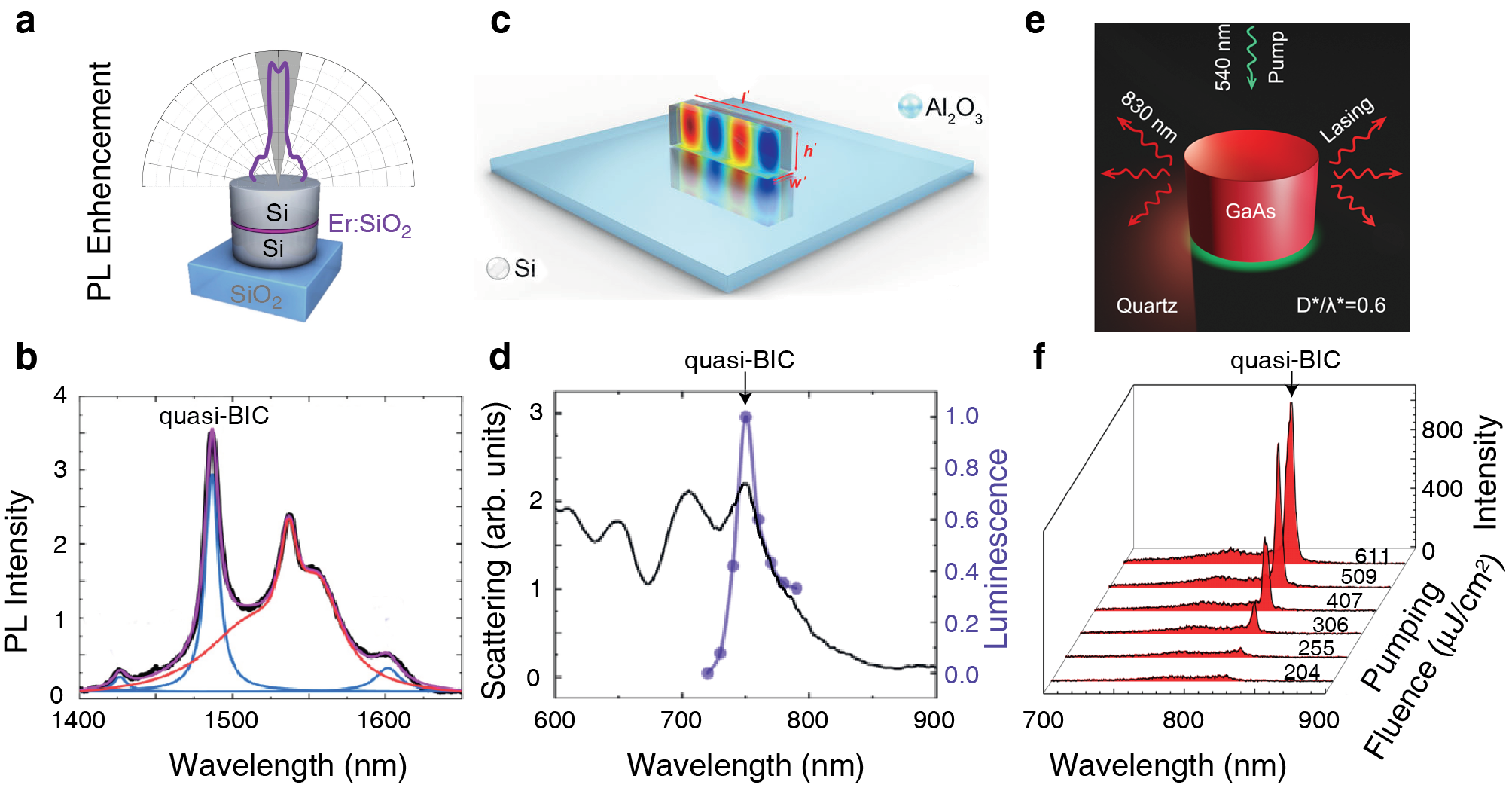}
\caption{\label{fig:emission}
\textbf{Emission from subwavelength nanoparticles enhanced by quasi-BIC modes.} \textbf{a} Schematic of the erbium-enriched silicon nanopillar and \textbf{b} its photoluminescence spectra (black curve) with a fitting (purple curve) decomposed into photoluminescence spectra of Er$^{3+}$ in SiO$_2$ (red curve) and peak associated with the nanopillar mode (blue curve). Disk diameter is 0.5$\lambda$. \textbf{c} quasi-BIC mode pattern in silicon cuboid and \textbf{d} its scattering (black curve) and excitation spectra (purple curve). Cuboid size is 0.4$\lambda$ by 0.4$\lambda$ by 1$\lambda$. \textbf{e} Schematic of a GaAs nanoscale cylinder for quasi-BIC driven lasing and \textbf{f}  evolution of the its emission spectrum at different pumping fluences. Disk diameter is 0.6$\lambda$. \textbf{a} and \textbf{b} are adapted from~\cite{kalinic2023quasi}; \textbf{c} and \textbf{d} are adapted from~\cite{panmai2022highly}; \textbf{e} and \textbf{f} are adapted from~\cite{mylnikov2020lasing}.
}
\end{figure*}

Recent review papers on Mie-resonant metaphotonics
has been just completed by Babicheva and Evlyukhin~\cite{vika} who focused on a comprehensive multipolar analysis, and discussed
a wide range of scattering phenomena that can be achieved within precisely engineered structures including the first and
second Kerker conditions and multipole engineering. 

At the end of this section, we wish making two comments. 
First, we notice that for some geometries higher-order Mie modes may have larger values of the $Q$ factor in comparison to the low-order modes. However, exploiting the low-order Mie modes provides a number of benefits including more straightforward ways for their excitation. 
Second, many effects discussed here are directly related to the Fano resonances associated with a weak coupling of high- and low-$Q$ modes~\cite{limonov2017fano}, retardation-type effects due to a finite-volume of the mode~\cite{rybin2013mie}, and even strong coupling case~\cite{rybin2017high}. Because of additional complexity of those effects, here for simplicity we consider 'Fano resonance' as an effect associated with an asymmetric spectral profile of the resonance~\cite{rybin2016purcell}.
Remarkably, for the scattering spectra of a dielectric cylinder, the Fano asymmetry parameter diverges at the parameters corresponding to the supercavity mode, as predicted theoretically~\cite{rybin2017high} and observed in experiment~\cite{melik2021fano}.

\section{Applications} 

The resonance particle modifies the optical effects mainly in three ways. First off, there is a narrow feature in the spectra, which bandwidth is related to the $Q$ factor. Noteworthy, the lineshape usually has asymmetric Fano profile~\cite{limonov2017fano}. The second way is the electric field distribution over the eigenmode volume, and the field enhancement is in direct relation to the $Q$ factor as well as the bandwidth. This leads to increasing light-matter interaction in hot spots, making the optical effects much stronger. The third one involves quantum nature. The dielectric particle perturbs the local density of photonic states, which are crucial parts of the radiation-related electronic transitions, and it is known as the Purcell effect. Theoretical studies uncover a relation between quasi-BIC mode and the enhancement of the Purcell factor~\cite{kolodny2019q,colom2022enhanced}.

Although sensing seems to be the most obvious application of high-$Q$ resonators, to date single particles supporting quasi-BIC modes were demonstrated for the microwave range only. The reason is that the analyte can be detected in the very vicinity of the particle, which makes such a sensor challenging in the optical range. On the other hand, the lower frequency is the bigger the volume of the object of interest. Shchelokova with coauthors~\cite{shchelokova2020ceramic} proposed to use a hollow cylindrical resonator made of high-index ceramic for efficient magnetic resonance imaging (MRI). The eigenmode has magnetic field enhancement inside the hollow, whereas the electric field is suppressed there. As a result a higher intensity of MRI signal is possible to collect for the same time, while the electromagnetic harm to the body is reduced. Another approach exploits the narrow band feature associated with the quasi-BIC modes supporting by the high-index cylinder~\cite{yusupov2021chipless, yusupov2023quasi}. The resonance spectral position dependence on permittivity can be detected with a better spectral resolution than those associated with lower-bandwidth regular Mie modes used in sensor devices. The ceramic cylinders were demonstrated to measure such parameters as environment temperature~\cite{yusupov2021chipless} or dielectric index of liquid flowing in a coiled around pipeline, which was designed to penetrate the quasi-BIC hot-spot surrounding the resonator~\cite{yusupov2023quasi}.

Modification of the optical emission rate is a fundamental property of photonic structures and radiation enhancement by a single nanoparticle supporting a supercavity mode is the most essential of their applications. 

Kalinic and coauthors~\cite{kalinic2023quasi} demonstrated the enhancement of the emission by matching the frequencies of a quasi-BIC mode in the dielectric disk and a quantum emitter. A thin oxide layer doped with erbium was placed inside a silicon nanopillar supporting quasi-BIC mode (Fig.~\ref{fig:emission}a). The erbium ion in a silica host is known as a light source with a sharp radiation peak at room temperature at telecom wavelength $\lambda=1540$ nm. The geometry of the nanopillar with the active nanogap was adjusted for the quasi-BIC mode to arise at $\lambda=1540$ nm (Fig.~\ref{fig:emission}b). As a result the nanogap modifies the dielectric environment of Er$^{3+}$ ions increasing its efficiency by the Purcell effect.  Such nanopillars were arranged in a square lattice, which leads to focusing 90\% of emission in a lobe normal to the sample with an angular width of about 10$^\circ$. The photoluminescence intensity was enhances by three orders of magnitude and decay rate of the Er$^{3+}$ emission was increased by 2 orders due to the Purcell effect.

Panmai and co-authors~\cite{panmai2022highly} considered silicon cuboid nanoparticles (see Fig.~\ref{fig:emission}c), which geometry was adjusted for a quasi-BIC mode to be formed as a mixed state of the ED, MD and MO modes. Although silicon is an indirect semiconductor with poor luminescence, the case of high-density carriers has different physics, the relaxation time of a hot electron from the $\Gamma$ point to the off-$\Gamma$ valley of the conductive band increases by two orders of magnitude, making possible a nonlinear luminescence process. Two-photon absorption of a femtosecond laser can be used as hot carrier excitation with efficiency proportional to the fourth power of the electric field inside silicon. The light-matter interaction causing the carrier generation at the quasi-BIC frequency is the most efficient (Fig.~\ref{fig:emission}d). Compared to bulk silicon, silicon cuboid nanoparticles demonstrated a significant improvement in quantum efficiency by six orders of magnitude to approximately 13\%.

Mylnikov with coauthors reported not just efficient photoluminescence but lasing action from subwavelength disk made of gallium arsenide (GaAs) at a cryogenic temperature~\cite{mylnikov2020lasing}.  They considered a supercavity mode originating from avoiding the crossing behavior of the TE$_{312}$ and TE$_{320}$ modes with the same azimuthal order 3. Because these two modes are predominantly coupled outside the nanoparticle, the Friedrich-Windgen approach predicts a redistribution of the radiation damping rate between two hybridized modes. The optimal GaAs nanodisk has a height of 344 nm and a diameter of 504 nm that results in a five-fold increase of the Q factor to almost 10$^3$ for the mode at 830 nm corresponding to the peak of GaAs emission at 77K (Fig.~\ref{fig:emission}e). However, realistic samples are placed on a quartz substrate and have remaining capping resist (hydrogen silsesquioxane) layer, which modify the mode structure. The measured Q factor of the nanodisks was 280. GaAs with remaining resist has a relatively broad photoluminescence spectrum at near-infrared wavelengths. The nanodisks were pumped by a femtosecond laser source at 539 nm. When the pumping fluence exceed 260 $\mu$J/cm$^2$ the narrow peak appears at 825 nm (Fig.~\ref{fig:emission}f) indicating the process of amplified spontaneous emission. The “S” sharper dependence on the pump fluence reveals that the nanodisk goes into the lasing action regime after a threshold of about 300 $\mu$J/cm$^2$.

\begin{figure*}
\includegraphics{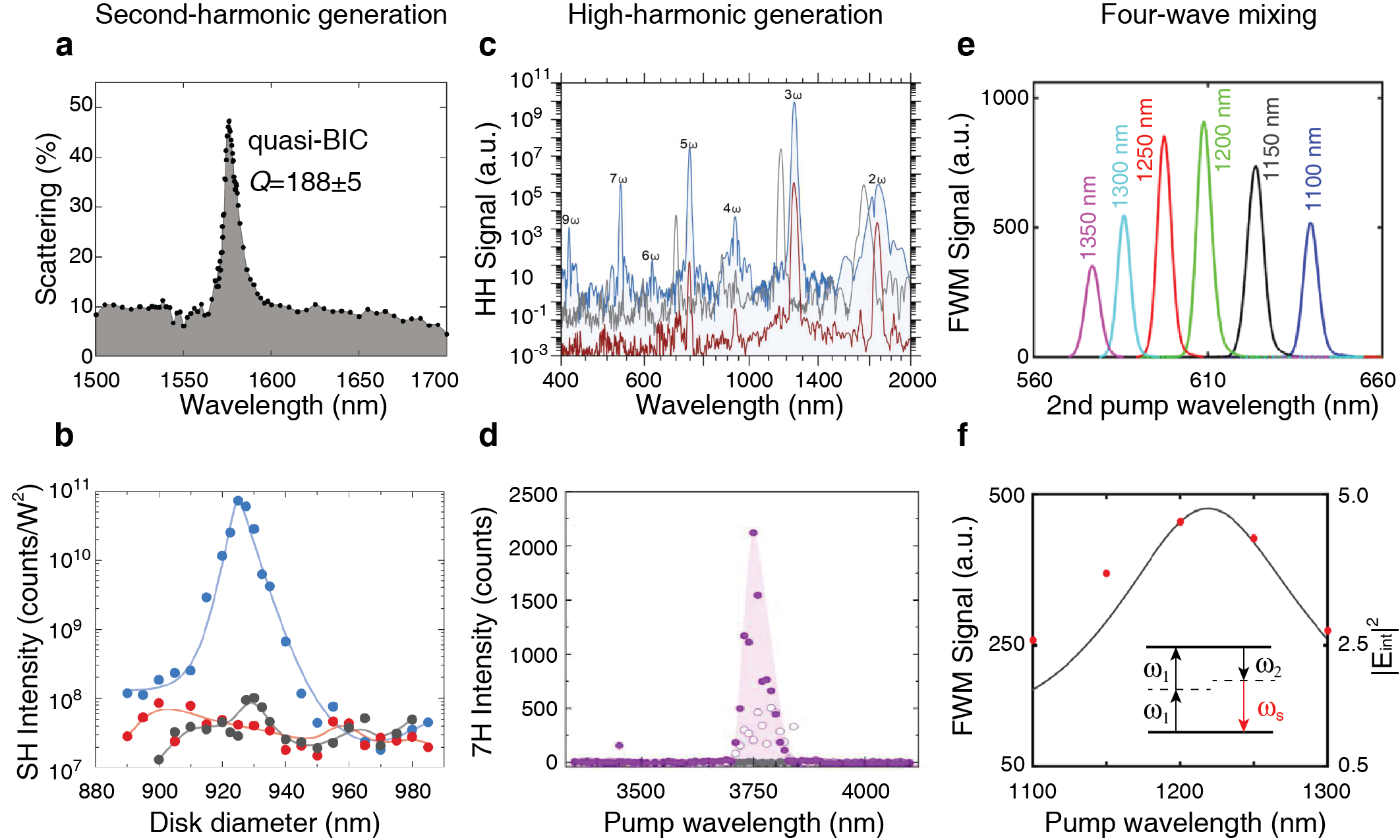}
\caption{\label{fig:harmonics}
\textbf{Nonlinear signal enchantment.} \textbf{a} Measured scattering spectrum of the optimal AlGaAs disk supporting the quasi-BIC mode and \textbf{b} second harmonic signal as a function of the diameter of the disk for different pump polarizations: linear (black), radial (red), azimuthal (blue). Disc radius is 0.3$\lambda$ at the fundamental frequency. \textbf{c} Theoretical spectra of optical harmonics from a subwavelength AlGaAs disk excited at resonance frequency 3750 nm (blue curve), off-resonance 3500 nm (gray curve) compared to the AlGaAs film (red). \textbf{d} Measured seventh harmonic intensity from the AlGaAs disk in dependence on the excitation wavelength. Solid symbols are azimuthally polarized excitation; open symbols are radial polarized excitation; gray dots, signal from unstructured AlGaAs film. Disc radius is 0.25$\lambda$ at the fundamental frequency. \textbf{e} Four-wave mixing signal measured from a silicon nanodisk when the wavelength of the second pump is changed (labeled at the peaks). \textbf{f} FWM signal normalized by the pump intensities (red dots) as a function of the wavelength of the second pump and the simulated spectrum of the electric field enhancement $<|\mathbf{E}_{inc}|^2>$ inside the nanodisk. Disc radius is 0.42$\lambda$ at the higher frequency. \textbf{a} and \textbf{b} are adapted from~\cite{koshelev2020subwavelength} ; \textbf{c} and \textbf{d} are adapted from~\cite{zalogina2023high}; \textbf{e} and \textbf{f} are adapted from~\cite{colom2019enhanced}.}
\end{figure*}

The localization of the electromagnetic field in a single nanoparticle supporting quasi-BIC enhances the efficiency of harmonic generation processes. Carletti with coauthors~\cite{carletti2018giant} consider nanocylinders with aspect ratio tuned to the condition for the quasi-BIC to appear. The nanodisk with permittivity 10.73 corresponding to AlGaAs at the telecom wavelength has the TE-polarized supercavity mode with Q=110, which has angular index $\pm1$. Taking into account $\chi^{(2)}=100 $pm/V, they study the second harmonic generation efficiency. In addition to a linearly polarized Gaussian beam with 30$^\circ$ incidence with respect to the cylinder axis, a focused azimuthally polarized be with index $\pm1$ was considered to better suit the supercavity mode. The strongest second-harmonic enhancement was found to be for the excitation of azimuthally polarized beam, and it is two orders of magnitude compared with the magnetic dipole mode in a similar nanocylinder. The multipolar nature of such supercavity modes in nanocylinders and its relation to illumination under structured light was studied by Volkovskya and co-authors~\cite{volkovskaya2020multipolar}. Koshelev with coauthors~\cite{koshelev2020subwavelength} reported experimental observation of the second harmonic generation from the nanodisk. They placed a subwavelength AlGaAs disk supporting a quasi-BIC at the telecommunication wavelength onto an ITO layer with a silica spacer amid. ITO layer additionally increases the Q factor to 188 (Fig.~\ref{fig:harmonics}a) because of the metallic properties in the frequency range, but it is transparent at the second harmonic frequency. Figure~\ref{fig:harmonics}b shows strong growth of the harmonic intensity when the disk diameter matches the supercavity aspect ratio. Also the plot compares that the effect is much stronger for the azimuthal polarization of pump illumination respective to the linear and radial (orthogonal) polarization.

High-harmonic generation from an AlGaAs disk was reported by Zalogina with coauthors~\cite{zalogina2023high} . They used mid-infrared pumping at wavelengths of 3.5-4.0$~\mu$m. Simulations show that the disk with dimensions 1384 nm (height) by 2050 nm (diameter) supports a supercavity mode with Q about 350, when placed at an aluminum oxide spacer of 700 nm for a separation with a 310 nm thick ITO layer. Experimental measurements reveals power scaling law with exponent 2.6 for fifth and 3.5 for seventh harmonics, which are two times smaller than predicted results with perturbative theory. In numerical simulations (Fig.~\ref{fig:harmonics}c) full-vector Maxwell equations were supplemented by nonlinear currents accounting for a dynamic Drude response of carriers excited to the conductive band. Similar to the case of second harmonic generation the polarization of the pump beam make a difference. In particular, Fig.~\ref{fig:harmonics}d shows the highest observed seventh harmonic for different illumination. For the azimuthal polarization the signal is five times of radial polarization of the illuminations.

High-$Q$ resonances in nanodisks can be used to enhance four-wave mixing that is a third-order nonlinear process. Colom with coauthors~\cite{colom2019enhanced} study the response of a silicon nanodisk with a height of 240 nm and diameter of 340 nm placed on a glass substrate. Two pump illumination does not require formation of a supercavity mode, while the system has the similar efficiency. The high-frequency pump wavelength $\lambda_1=810$ nm matches the electric dipole resonance, while the low-frequency pump wavelength $\lambda_2$ was swept over the magnetic dipole resonance at about 1200 nm. The four-wave mixing signal is shown in Fig.~\ref{fig:harmonics}e. The signal intensity follows the electric field enhancement in the silicon nanodisk (Fig.~\ref{fig:harmonics}f). Although when no hybridization of two resonances takes place, the conversion efficiency is similar to one of a supercavity mode, which strongly relaxes the tolerance of the nanodisk geometry, however the process requires independent adjustment of two laser pulses not only in the wavelengths but also in space and time. We notice that for many nonlinear effects with isolated nanostructures such as harmonic generation, the mode overlap inside a resonator plays a crucial role for enhancing efficiencies of nonlinear processes. 

\begin{figure*}
\includegraphics{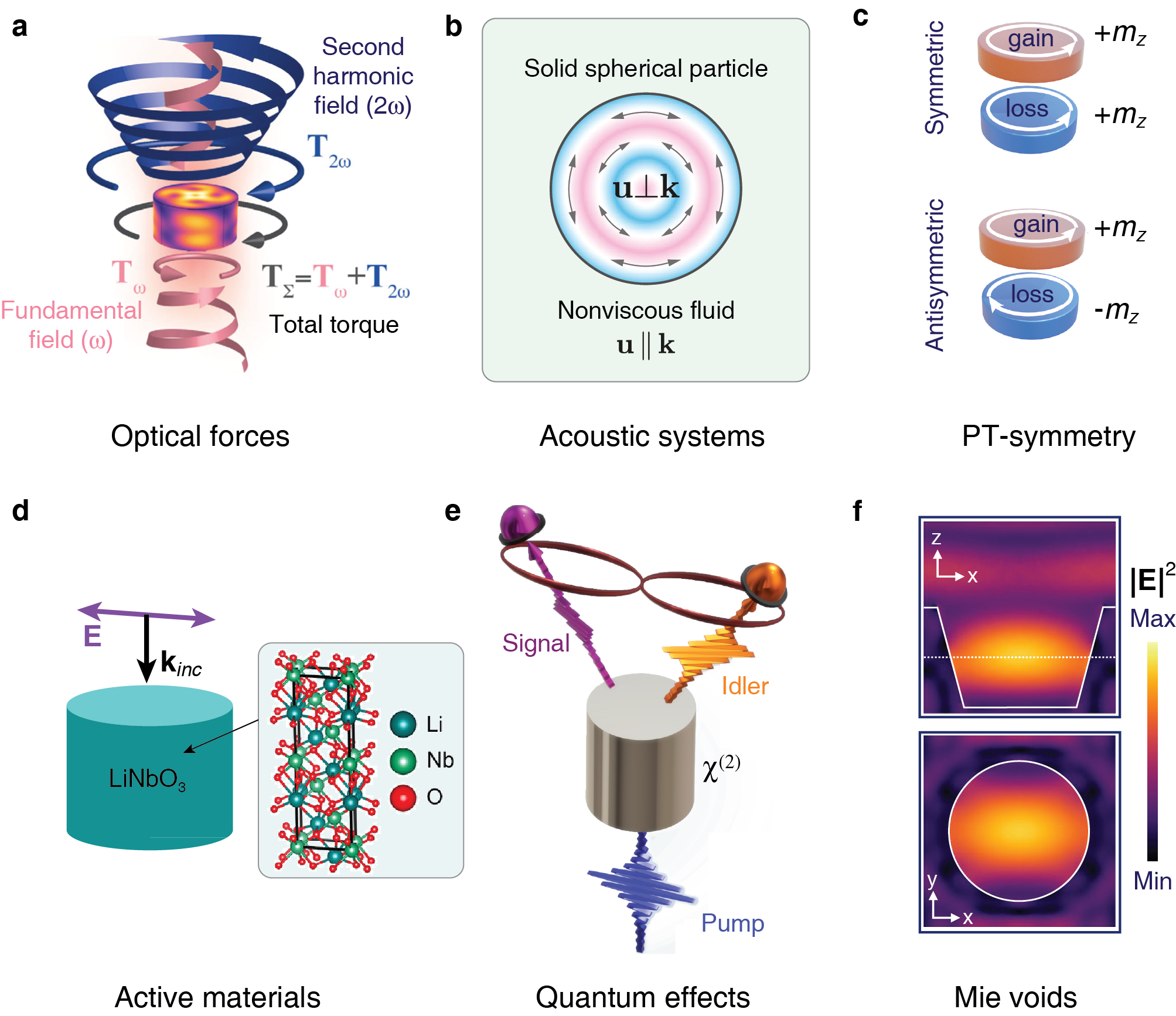}
\caption{\label{fig:neweffects}
\textbf{Emerged directions in single-particle quasi-BIC.} \textbf{a} Nonlinearity-induced optical torque arising due to different angular momentum of the fundamental and second harmonics. \textbf{b} Acoustic transverse BIC mode supported by solid spherical particle immersed into nonviscous fluid supporting longitudinal waves only. \textbf{c} 
A two-layer metasurface comprising two types of dielectric disks, which supports BIC and EP. \textbf{d} Pumping configuration of LiNbO$_3$ nanopillar for efficient optical-to-THz conversion process. \textbf{e} Photon pair-generation in dielectric resonance cylinder;  the entangled signal and idler photons can be generated and detected in the far-field. \textbf{f} Mie modes in dielectric spheres and Mie modes in air void in the dielectric host.
\textbf{a} is adapted from \cite{toftul2023nonlinearity}; \textbf{b} is adapted from~\cite{deriy2022bound}; \textbf{c} is adapted from~\cite{valero2023exceptional}; d is adapted from~\cite{arregui2023thz}; \textbf{e} is adapted from~\cite{weissflog2023nonlinear}; \textbf{f} is adapted from \cite{hentschel2023dielectric}.
}
\end{figure*}

\section{Generalisations}

Subwavelength objects supporting quasi-BIC in a form of supercavity modes provide many opportunities for metaphotonics. Although many exciting phenomena were reported much more effects have yet to be reported. The first issue is a challenge to scaling down the effects. In particular, we discussed above the sensor devices operating in the microwave frequency range. The nanophotonic subwavelength sensing has high spatial precision, which is an advantage but requiring high values of experimental tolerance simultaneously. Other effects become important only in the optical range. Optical forces are one of them. Toftul with coauthors~\cite{toftul2023nonlinearity} proposed to induce optical torque in the axial symmetric disk through the generation of the second harmonic due to the different angular momentum of the fundamental and second harmonics (Fig.~\ref{fig:neweffects}a). A nanodisk with GaAs parameters supports two supercavity modes at the fundamental frequency with angular momentum 4 and a mode at the second harmonic frequency with angular moment 1. Q factor of both modes are about 40, which make the torque induction reasonable.

Novel physics might appear with a change of wave nature. Deriy with coauthors~\cite{deriy2022bound} considered acoustic resonator of spherical shape. In contrast to the electromagnetic, acoustic waves are known to be two types which are pressure waves with longitudinal polarization and shear waves with transversal polarization and different materials support different types of acoustic oscillations. Solids allow both types of waves, while nonviscous liquids only longitudinal waves are possible. So the Mie-type transversal modes in solid spherical particle are orthogonal with the longitudinal spherical waves in the surrounding fluid because of the symmetry (Fig.~\ref{fig:neweffects}b). However even a weak deformation of the spherical shape results to appearance of the quasi-BIC acoustic modes. 
Laypina \emph{et al}.~\cite{lyapina2015bound}, and later Huang \emph{et al}.~\cite{huang2021sound,huang2022general}, analyzed an acoustic resonator coupled to a waveguide supporting a continuum of radiative modes. By tailoring the resonator geometry, it makes possible in such a system to tune a pair of acoustic modes to hybridize them into a Friedrich-Wintgen quasi-BIC state. The effect was studied theoretically and verified experimentally. 

Randerson \emph{et al}.~\cite{randerson2024high} studied WS$_2$ nanodisks deposited on a gold substrate. The nanodisk supports Mie modes, and the substrate provides a medium supporting surface plasmon polaritons. By adjusting the disk radius, the authors tuned the mode frequencies achieving a mode hybridization with the formation of a supercavity mode. The experimentally measured $Q$ factor of the supercavity mode was up to 33 times higher of each non-interacting Mie mode. Theoretical studies revealed that a thin hBN spacer placed between the nanodisk and the gold substrate causes a strong enhancement of the electric field and increasing the Purcell factor.

Parity-time (PT) symmetric particles described with non-Hermitian Hamiltonians still have real spectrum until a parameter does not reach the critical value at which two real eigenfrequency coalescence and become degenerate while decay rate appears for both modes. The coalescent point is referred as to exceptional points, at which the eigenfunctions degenerate and the Hamiltonian corresponds to a defective matrix. In photonics PT-symmetric systems are realized with loss and gain media swapped their locations under inversion of the coordinate. In electromagnetic problem radiation losses confuse the idealistic picture of PT-symmetry, but the exceptional-point feature survives. Valero {\it et al.}~\cite{valero2023exceptional} designed a structure merging BIC and exceptional points (Fig.~\ref{fig:neweffects}c), which does not radiate due to BIC and has a square root sensitivity to a perturbation inherited from exceptional points.

Leon with coauthors~\cite{arregui2023thz} proposed to use the resonances in nanodiscs to generate terahertz radiation through optical rectification, which is a second-order nonlinear process consisting in generation of terahertz photons at the frequency difference between two optical pulses at about 1.5$\mu$m. They consider disk made of material with properties of AlGaAs and lithium niobate having strong second order susceptibility tensors (Fig.~\ref{fig:neweffects}d). The resonance condition for the optical frequencies leads to efficient optical rectification process. 

Another second-order nonlinear process enhanced with resonances in single subwavelength nanodisks was studied by Waissflog {\it et al.}~\cite{weissflog2023nonlinear}. Process of spontaneous parametric down-conversion results in generation of entangled Bell states with a pair of photons (Fig.~\ref{fig:neweffects}e). They consider illumination of the nanodisk made with high second-order susceptibility tensor typical for materials with zinc-blind crystal lattice such as AlGaAs. As for other nonlinear-based effects the pump wavelength was tuned to the resonance in the nanodisk, which results in entangled signal and idler photon generation efficiency.

Recently, Hentschel {\it et al.}~\cite{hentschel2023dielectric} reported on Mie-type resonances excited in voids in a silicon host matrix (Fig.~\ref{fig:neweffects}f). In general, the Mie theory is applicable for any permittivity values of the sphere and surrounding media. Usually, the wave tends to locate in high-index materials so the Mie resonances in air spheres have low-$Q$ factors leading to form in the scattering a slow-modulated background spectrum. However, silicon in visible spectral range has strong interaction with electromagnetic field, and it acts as a mirror, so that the light waves are confined in the air voids producing narrow resonances. Interestingly enough, the physics of Mie voids can be described through the generalized quasi-Babinet principle~\cite{hamidi2023quasi}. The Mie voids can be employed for a design of structured colors~\cite{opn_2024} and other applications.

\section{Summary and outlook}

We have discussed the recent trends in the study of Mie-resonant dielectric subwavelength structures that can be employed for many applications in nanophotonics, including enhanced light-matter interactions, strong nonlinear effects, highly efficient sensing, and low-power nanoscale lasers. 

There are several ways to improve the functionalities of single-particle subwavelength devices. First, this is the development of more accurate and stable nanofabrication techniques to sustain the resonator geometry. Second, the studies of other geometries such as conical resonators and other types of non-conventional shapes allowing 
to bring many features of non-Hermitian physics. Last but not least, 
high-$Q$ nanoparticles can be fabricated from functional materials enabling active effects and applications, as discussed below. 


Many novel ideas in Mie-tronics have been suggested more recently. In particular, local density of photonic states has been studied for nanodisks~\cite{mignuzzi2019nanoscale}, and thermorefractive bistability has been discussed for dielectric particles~\cite{ryabov2022nonlinear}. 
Enhancement of a strong coupling between the Mie modes and excitons in {\it transition metal dichalconedite structures} (or TMDC) has been demonstrated in many theoretical and experimental studies starting from the pioneering demonstration of a coupling between an anapole and excitons in WS$_2$ nanodisks~\cite{Shegai_2019}. Raman nanolaser driven by quasi-BIC resonances in a nanodisk has been proposed in Ref.~\cite{riabov2023subwavelength}, and the concept of interfering resonances and Mie modes in dielectric nanodisks has inspired novel research on similar quantum states~\cite{nefedkin2021quantum}.


Meta-atoms are the building blocks of Mie-tronics, and 
usually they are simple dielectric resonators. However,
more complex meta-atoms are also possible e.g. those in the form of {\it supercrystal meta-atoms} composed of coupled perovskite quantum dots~\cite{pavel}. The multiscale structures exhibit specific emission properties, such as spectrum splitting and polaritonic effects, and they can provide novel functionalities in the design of many novel types of active metasurfaces.


Complexity of meta-atoms can be already realized with much simpler {\it core-shell geometry}.  The functionalities of a Mie-resonant dielectric nanoparticles can be greatly extended by decorating the surface with various passive and active materials~\cite{wei}. Recently, we have witnessed the development of advanced fabrication processes of the core/shell architectures for a variety of shell materials that modify the properties of silicon nanoparticles and introduce new functions. The shell materials include passive low-refractive index materials, materials of tunable optical properties, fluorescence dyes, transition metal dichalcogenides, and noble metals with surface plasmon resonances~\cite{core_shell}. 


One more important direction is to study the resonant interaction between different types of Mie modes and guided-wave modes in extended structures. One such example is the interaction with {\it epsilon-near-zero (ENZ) modes}~\cite{boyd_2023}. Analytical, simulation, and experimental analyses reveal that the presence of the ENZ substrate significantly modifies the electric and magnetic Mie dipole modes, when coupled to the ENZ mode, indicating strong coupling with notably large subpicosecond nonlinear responses. 


Last but not least, very important recent developments in Mie-tronics are associated with the application of {\it machine learning methods} for multi-parameter optimizations and inverse design strategies to achieve the specific optical properties of Mie-resonant nanoparticles. Recently, the team of M. Litchinitser~\cite{natasha_2024} demonstrated physics-empowered forward and inverse machine-learning methods to design dielectric meta-atoms with controlled multipolar responses. By utilizing the multipole expansion theory, their model efficiently predicts the scattering response of meta-atoms with diverse shapes and desired multipole resonances.  Implementing the inverse design model, one can predict the wavelength-dependent electric field distribution inside and near the meta-atom, and uncovering new regimes of light-matter interaction.


We should notice that in all examples above, we have restricted our examples by the cases of isolated subwavelength resonators. Many more interesting effects are expected in {\it pairs of Mie resonators}~\cite{liza,dmitriev2019combining,dmitriev2021optical} or {\it arrays of identical Mie resonators}~\cite{hoang}.  By employing the multipole Mie scattering, one can demonstrate the emergence of photonic flatbands due to fine-tuning of interaction in a one-dimensional
chain of high-index dielectric nanoparticles~\cite{hoang}. 
In addition, multiple scattering theory allows to study coupling between different multipoles beyond the short-range approximation~\cite{hoang}. 


We believe that the concepts of Mie-resonant photonics (or Mie-tronics) will be influential for other fields such as photovoltaics, optical imaging, polaritonics, as well as quantum technologies. Tailored resonances of high-index dielectric subwavelength structures and metasurfaces can boost nonlinear response of hybrid materials being combined with two-dimensional materials creating flexible and tunable optoelectronic metadevices. Importantly, active Mie-resonant nanoantennas can be employed as the smallest light sources for dense photonic integration of on-chip metadevices. Combining the advantages of all-dielectric metasurfaces with a high-$Q$ resonances would allow to achieve tunable control over the electromagnetic fields, also realizing novel types of chiral biosensors, and thus increasing both device sensitivity and their multiplexing abilities. We believe that modern integrated photonics requires the developments in device design, material synthesis, nanofabrication, and characterization, and the combination of all those efforts can be realized with all-dielectric metaphotonics. We anticipate many novel discoveries in Mie-tronics to be demonstrated in the coming years.

\section{Acknowledgments}

The authors thank their numerous colleagues and collaborators from the Australian National University in Canberra and the ITMO University as well as the Ioffe Institute in St.~Petersburg for productive collaboration and useful discussions. They thank also Victoria Babicheva, Andrey Bogdanov, Andrei Evlyukhin, and Wei Liu for their critical reading of an initial version of this manuscript and useful suggestions, and Baohua Jia for her kind invitation to contribute a perspective paper to this new journal. 

Y.K. acknowledges a financial support from the Strategic Fund of the Australian National University, the Australian Research Council (grant DP210101292), and the International Technology Center
Indo-Pacific (ITC IPAC) via Army Research Office (contract
FA520923C0023). M.R. acknowledges a financial support from the Russian Scientific Foundation (grant 24-72-10038). 

\section{Contributions}
M.R. and Y.K. wrote the manuscript text and prepared figures. All authors reviewed the manuscript.

\section{Competing interests}
The authors declare no competing interests.


\end{document}